\documentclass[]{elsart}
\usepackage{graphicx}   
\usepackage{natbib}
\begin{document}
\begin{frontmatter}

\title{Modulation of the reaction rate of regulating 
protein induces large morphological and motional change of amoebic cell}

\author[label1]{Shin I Nishimura} 
\ead{shin@tbp.cse.nagoya-u.ac.jp}
\author[label1,label2]{and Masaki Sasai}

\address[label1]{Department of Computational Science and Engineering, Graduate School of Enginieering, 
Nagoya University Nagoya 464-8603, Japan}
\address[label2]{CREST-Japan Science and Technology Agency, 464-8603, Japan}

\begin{abstract}
Morphologies of moving amoebae are categorized into two types.  
One is the ``neutrophil'' type in which the long 
axis of cell roughly coincides with its moving direction. 
This type of cell extends a leading edge at the front and retracts 
a narrow tail at the rear, whose shape has been often drawn as a typical  
amoeba in textbooks.  The other one is the ``keratocyte'' type with 
widespread lamellipodia along the front side arc. 
Short axis of cell in this type roughly 
coincides with its moving direction.
In order to understand what kind of molecular feature causes conversion 
between two types of morphologies, and how two typical morphologies are maintained,
a mathematical model of amoebic cells is developed. This model describes movement of cell
and intracellular reactions of activator, inhibitor and actin filaments in a unified way.  
It is found that the producing rate of activator is a key 
factor of conversion between two types. 
This model also explains the observed data that the keratocye type cells tend to 
rapidly move along a straight line.
The neutrophil type cells move along a straight line when the moving velocity is small, but
they show fluctuated motions deviating from a line when they move as fast as the keratocye type cells. 
Efficient energy consumption in the neutrophil type cells is predicted.
\end{abstract}

\end{frontmatter}

\section{Introduction}
Amoebic cells have been used as model systems to analyze
the molecular mechanism of cell locomotion \citep{Pollard03a}.
Chemical or mechanical signals arriving on amoebic cell membrane 
induce the intracellular signaling cascade, resulting in the activation of Arp2/3 complex to
initiate actin polymerization at the front edge of cell \citep{pollard03b}. 
Thus elongated actin filament drives protrusion at the cell anterior. 
Myosin II seems to play roles in the retraction process at the cell posterior \citep{Eddy00}, 
and the cell moves forward when protrusion at the front is 
followed by retraction at the rear.
Although those molecular details have begun to be elucidated,  
there still remain large mysteries on how biochemical reactions drive the
coordinated change of morphology.

Morphologies of moving amoebae seem to be categorized into 
two types. One is the ``neutrophil'' type in which the long 
axis of cell roughly coincides with its moving direction. 
This type of cell extends a leading edge at the front and retracts 
a narrow tail, whose shape has been often drawn as a typical  
amoeba in textbooks.  The other one is the ``keratocyte'' type with 
widespread lamellipodia along the front side arc. 
Short axis of cell in this type roughly 
coincides with its moving direction. Fish or amphibian epidermal keratocytes 
keep their stable half-moon shape during locomotion and rapidly crawl along the
relatively straight line. This stability in morphology and motility 
has allowed detailed investigations on keratocytes \citep{Lee93}.
Although there is further variety in cell shapes depending 
on the strength and frequency of protrusion and structures of cytoskeltons, 
we here focus on the difference whether the shape is head-tail long or 
laterally long and call the former ``neutrophil type '' and the latter ``keratocyte type ''.

Interesting observations are that changes in a single or a small number of genes
can convert cells from the neutrophil type to the keratocyte type: 
Breast adenocarcinoma cells overexpressing a Cas-family protein showed morphological
conversion to the keratocyte type \citep{Fashena02}. A cell line of human
microvascular endothelial cell moved in a keratocyte-like manner \citep{Kiosses99, Fischer03}.
An interesting example of conversion is $amiB$-Null {\it Dictyostelium} cells whose 
{\it amiB} gene is disrupted.  
The wild type {\it Dictyostelium} cells are representative neutrophil type cells and move
toward the aggregation center with repeated dynamical protrusion and retraction when they are starved.
The starved {\it amiB}$^{-}$ cells, on the other hand, 
move unidirectionally with maintaining a half-moon shape \citep{Asano04}. 
A role of 
AmiB 
in motility has not yet been known but this example clearly showed that
the change in a single gene caused distinct conversion of morphology from the neutrophil type 
to the keratocyte type.

Here, in this paper we ask questions of what kind of molecular feature causes conversion 
between two types of morphologies, and how two typical morphologies are maintained.
In order to answer these questions, theoretical model is necessary to be developed, which can 
treat motion, sensing, and chemical reactions in a unified way.
Such mathematical model of amoeba has been developed by
\cite{NishimuraAlife, NishimuraArob, NishimuraPRE}, and it has been shown that the
time lag between the shape change and the intracellular molecular diffusion induces cell
locomotion with ``inertia"-like features: The model can explain experimental observations that
once cells are boosted in one direction, cells can keep moving forward without 
chemoattractants \citep{Verkhovsky99} or even toward the direction 
of decreasing chemoattractants \citep{Jeon02}. It was predicted that in 
a ``centripetal" distribution of chemoattractants cells can show a rotatory motion by 
keeping fictitious angular momentum \citep{NishimuraPRE}. 
These fictitious inertia or angular momentum results from the
collective dynamics of morphological change and chemical reactions and diffusion.
In the present paper we use this model to explain the dramatic change of morphology between
the neutrophil type and the keratocyte type.

\section{Model} 

As in fluid or solid mechanics, the movement of an extended object 
can be modeled with either way of descriptions, the Lagrange description or the Euler 
description. In the Lagrange description a cell is regarded to be composed of many small pieces 
and equations of motions of these pieces are considered. 
What is advantageous in the Lagrange description is the easiness to
treat forces and the mechanical balance among them in cell. 
In the Euler description, on the other hand, the movement of cell is described from the coordinate 
fixed in space. This description is useful to treat chemical distributions and reactions in cell. 
In models of amoebae  both two descriptions were adopted
in a hybrid manner  \citep{Bottino01,Rubinstein05} to treat the force balance and 
the chemical reactions at the same time.
Another way to express amoebae is to solely use the Euler description.
In this case phenomenological rules are used instead of treating 
the force balance in equations of motions \citep{Satyanarayana04}.
The advantage of this approach is the simplicity in modeling, 
which allows highly efficient simulation of cell movement.
We here adopt the Euler description to calculate a large number of 
trajectories for statistical sampling. 

We introduce hexagonal grids that indicate the simulation space.  
A cellular body is thought of as a set of single connected grids 
in the grid space.  We call these grids ``cellular'' grids. 
We call remaining grids in the model space ``external'' grids. 
It has been known that the moving cell repeats adhesion and de-adhesion to 
substratum, suggesting that the cell movement is regulated by a ``clutch-like'' 
mechanism \citep{Smilenov99}. 
Here, we simply assume that the cellular grids 
are linked to the substratum by adhesive molecules and do not slide. 
Although there are complex biochemical pathways to activate and inactivate 
adhesive molecules such as integrins and 
associated proteins, we omit modeling these processes for simplicity. 
In our model conversion of the external grid to the cellular grid and conversion
of the cellular grid to the external grid are assumed to represent 
adhesion and de-adhesion processes, respectively. 
If a cellular grid connects at least one external grid, we 
call this grid  a ``membrane'' grid. If a cellular grid 
is not a membrane grid, we call that grid ``cytoplasm'' grid. 
There are three different molecular densities defined in each 
cellular grid: activator, $A$, inhibitor, $I$, and actin filaments, $F$.  
In addition, concentration of the external chemical signal accepted by receptors, $S$,
is defined at each membrane grid. 
Activator and inhibitor are regulating proteins. Though the regulating biochemical network
has not yet been fully identified, PI3K and PTEN are candidates of proteins
working as activator and inhibitor, respectively \citep{Levchenko02}. 
It is known that PI3K phosphorylates PI(4,5)P$_2$ to PI(3,4,5)P$_3$, 
whereas PTEN dephosphrylates PI(3,4,5)P$_3$ to PI(4,5)P$_2$, working in an antagonistic way \citep{Iijima02}.

We use ``rule based dynamics'' in which several rules are called randomly. 
The following paragraphs explain those rules.

(1) {\bf Kinetics}: Both activator and inhibitor are 
produced by the stimulation of the external signal \citep{Levchenko02}.  
The activator enhances polymerization of actins, whereas 
the inhibitor suppresses the polymerization. First, this rule selects 
a grid in the cellular domain randomly. If densities of activator, 
inhibitor and actin filaments at the selected grid $j$ are expressed 
as $A_j$, $I_j$ and $F_j$, respectively, those variables are 
changed obeying the following equations:
\begin{eqnarray}
  {A'_j} &=&  A_j + \alpha  S_j - k_{\alpha} A_j \\
  {I'_j} &=&  I_j + \beta   S_j - k_{\beta}  I_j  \\
  {F'_j} &=&  F_j+\left\{
\begin{array}{ll}
  \gamma -k_f F_j & (\frac{A_j}{I_j}>h )\\
  -k_f F_j            & (\mbox{otherwise})
\end{array}
\right.,
\end{eqnarray}
where $\alpha$, $\beta$, $\gamma$, $k_{\alpha}$, $k_{\beta}$,
$k_f$,  $\gamma$ and  $h$  are constants. 
$\alpha$ and $\beta$ are the creation rates of activator and inhibitor, respectively, 
and $\gamma$ is the rate of polymerization of actin filaments. 
$k_{\alpha}$ and $k_{\beta}$ are the degradation rates of activator and inhibitor, 
and $k_{f}$ is the de-polymerization rate of actin filaments. 
Increase in $h$ decreases the possibility of polymerization of actin filaments. 
$S_j$ indicates  
the concentration of chemoattractants or the strength of the 
external signal at the $j$th grid. 
$S_j$ is set to zero if the $j$th grid is in cytoplasm.

(2) {\bf Diffusion}: Only the inhibitor diffuses into the 
whole cytoplasm \citep{Levchenko02}. This rule selects a grid 
from the whole cellular domain. At the selected $j$th grid 
and its nearest cellular $l$th grid, $I_j$ and $I_l$ obey the following equations:
\begin{eqnarray}
\label{diffusion1}
  I'_j &=& I_j - DI_j  \\   
  I'_l &=& I_l + \frac{DI_j}{n}, \label{diffusion2}
\end{eqnarray} 
where $D$ is the diffusion constant. $n$ is the number of 
the nearest cellular grids. $D$ should be smaller than 1 by definition.
Since activator and actin filaments do not diffuse and the external 
signal $S_j$ is nonzero only at membrane grids,
activator and actin filaments distribute around membrane grids. 

(3) {\bf Keeping the cell}: We also give a rule to prevent 
cell from breaking into pieces. The cellular volume is kept 
and the cellular surface length is constrained to be as small 
as possible. This rule randomly selects a grid from the membrane. 
Then the rule decides either to remove the grid or to 
add a new cellular grid around the grid. 
This rule checks the cellular tension by calculating a cost function, $E$, 
as $E = (V-V_0)^2+cL^2$, where $V$ is the total number of cellular grids and $L$ is 
the total number of membrane grids, and $V_0$ and $c$ are constants. 
$V_0$ is the equilibrium number of cellular grids and $c$  is a "stiffness"-like factor. 
When $E'$ denotes the cost function after a cellular grid is removed or added, 
we define the probability $P$ as
$P=\exp\left(-\frac{E'-E}{kT}\right)$, 
where $kT$ is a constant 
to control the extent of fluctuation. 
We generate a random number 
between 0 and 1 and then compare the number with $P$. 
If the number is smaller than $P$, we ``undo'' the event 
of removing/adding. 
Note that if ``removing'' is chosen, the values of $A$, $I$ and $F$ 
in the removed grid are added into the nearest cellular grid. 
If ``adding'' is chosen, the three values are inherited from a 
randomly selected nearest grids, and three values of 
the selected nearest grid are reset to zero. 

This rule is related to contraction of the rear of the cell. 
Recent experimental studies have elucidated that interactions among 
myosin II and actin filaments generate forces to contract the rear.
For cellular locomotion, protrusion at the leading edge seems to have a prime role of 
cellular ``decision making'' about where the cell should go \citep{Weber06}, and
the myosin-generated contraction works to weaken the load of protrusion at 
the leading edge. In this way the rate of contraction may be regulated 
so that the cell size is kept as constant as possible. 
We should note that not only the myosin II contraction 
but also the cellular cortical tension exerted by actin-myosin network is a factor of keeping the cell size,  
where myosin I seems to contribute to the generation of cortical tension \citep{Dai99}. 
Thus, the effects of the rear contraction and the cortial tension could be modeled by imposing the rule of
keeping the cell size.
Several theoretical studies on cell locomotion explicitly or implicitly assumed 
the regulation of keeping cell size in their models \citep{Bottino01,Rubinstein05}.
Here, we also impose the rule of keeping cell size instead of considering the
detailed mechanisms of contraction and cortical tension.  
This simplification might be partly justified by the experimental data that 
the shapes of wildtype {\it Dityostelium discoideum} does 
not differ much from those of 
{\it myosin-II}-null 
mutants even though 
the exerted mechanical forces are clearly different \citep{Uchida03}.  
The {\it amiB} and {\it myosin II} double nockout mutants are also known to show a keratocyte-like shape
as in mutants which only lacks AmiB \citep{Asano04}.  

(4) {\bf Cellular domain extension}:
The rule randomly selects a grid from the membrane. 
When $F_j$ at the selected $j$th grid in the membrane exceeds 
the threshold $F_{th}$, an external grid in the six nearest 
grids of the $j$th grid is turned into a cellular grid. 
$F_{th}$ roughly defines the amounts of  actin filaments per a grid.  
When there are two or more than two external grids 
around the $j$th grid, a grid is randomly selected.  
If this grid is referred to as $l$, $F_l=F_j/2$ and 
other variables are set to zero. $F'_j$ equals to $F_j/2$ by definition, 
where the prime indicates the value at the next time step.

(5) {\bf Sampling}:
The event of sampling does not alter the system but 
the cellular shape, position, and concentrations of molecules are monitored.  
``One step'' of the cellular dynamics is counted when Sampling is called once.

We also give a master rule that randomly selects 
one of the above rules. Probabilities of selection for rules from (1) to (5) 
are written as $P_1$, $P_2$, $P_3$,$P_4$ and $P_5$, respectively with
$P_1+P_2+P_3+P_4+P_5=1$. 
When the master rule selects one of the five rules, 
the selected rule is executed. We iterate this process several millions times.

We assume that the length of a grid is approximately 1 $\mu$m, 
we put the initial shape of the cell to be a circle with  
30-grid diameter, and the equilibrium volume is set to be 
$V_0=900$. The effective diffusion constant of the inhibitor is $D_{\mbox{eff}} \equiv 
(D/6) (1.0 \mu \mbox{m})^2/\delta t \times (P_2/P_5)/V_0$, 
where $\delta t$ is the time length of one step. 
By setting $\delta t=0.3$s, $P_2=0.8718$, $P_5=0.00029$, and $D=0.45$, 
we have $D_{\mbox{eff}}\approx 0.8 \mu \mbox{m}^2/s$, 
which is of the same order of the diffusion constant of membrane proteins \citep{Gamba2005}. 
Other parameters are set to prevent the actin filament 
from spreading too broadly along the membrane but 
to be heterogeneously distributed in response to the anisotropic 
environmental stimuli \citep{Iijima02}; $\beta=0.1$, 
$k_{\alpha}=0.9$, $k_{\beta}=0.02$, $\gamma=4.0$, 
$k_f=0.99$, $h=10.0$, $F_{th}=1.0$, $P_1=0.040686$, 
$P_3=0.05812$, $P_4=0.02906$, $c=1.2$ and $kT=20$.

The cell locomotion is simulated with
two different patterns of the external 
signal distributions $S_j$: One is ``linear gradient'' 
$S_j= 0.0222y_j+1$, where $y_j$ indicates the position of the $j$th grid
along the $y$ direction in the grid space,
and the other is the uniform distribution $S_j=1.0$.  
The cell locomotion follows the gradient of the signal distribution in the former case.
Even in the uniform signal distribution of the latter case 
spontaneous polarization and unidirectional
movement of cell have been experimentally observed \citep{Wilkinson98}. 
In the present model intracellular fluctuations of activator and inhibitor
are amplified by the fluctuation of the cell shape and grow into 
the spontaneous cellular locomotion, for which we make observation 
on the relation between motility and morphology.

\section{Results}

\begin{figure}[p]
\begin{center}
\includegraphics[scale=0.7]{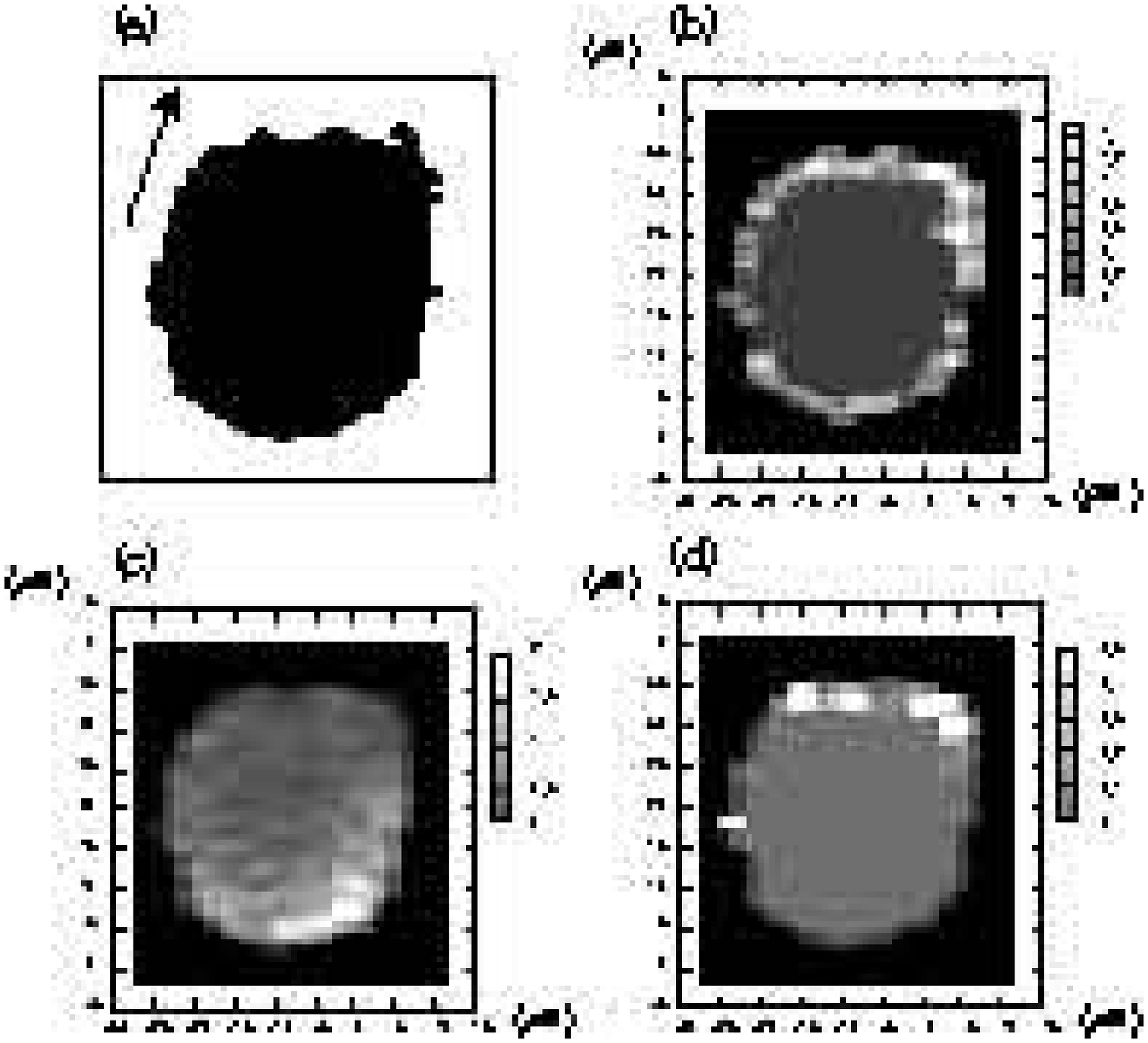}
\caption{(a) A snapshot of the cell shape with $\alpha=1.0$. 
An arrow indicates the direction of cellular motion (b) 
Distribution of activator in the same cell as in (a). 
Local densities are expressed with gray scale. 
(c) Distribution of inhibitor. (d) Distribution of actin filaments. }
\label{SNAP1}
\end{center}
\end{figure}

When the parameter $\alpha$, the rate of activator creation, 
is set to 1, the long axis of the cell shape roughly 
coincides with the moving direction of the cell (Fig. \ref{SNAP1}(a)).
The external signal is distributed to have the linear gradient. 
Fig.\ref{SNAP1}(b), (c) and (d) indicate localization of activator, 
inhibitor and actin filaments, respectively. 

Activator accumulating at the membrane is more localized around 
the front of cell than the rear, whereas inhibitor 
distribution is biased towards the rear. As a consequence of such localization of
activator and inhibitor, actin filaments are created at the front of the cell. 
These localized patterns are similar to those observed in {\it Dictyostelium discodeum}  \citep{Iijima02}, 
where PI3K and F-actin are biased towards the front, whereas PTEN accumulates at the  rear of {\it Dictyostelium}, 
showing that the cell with $\alpha=1$ is the neutrophil type.  
 
\begin{figure}[p]
\begin{center}
\includegraphics[scale=0.7]{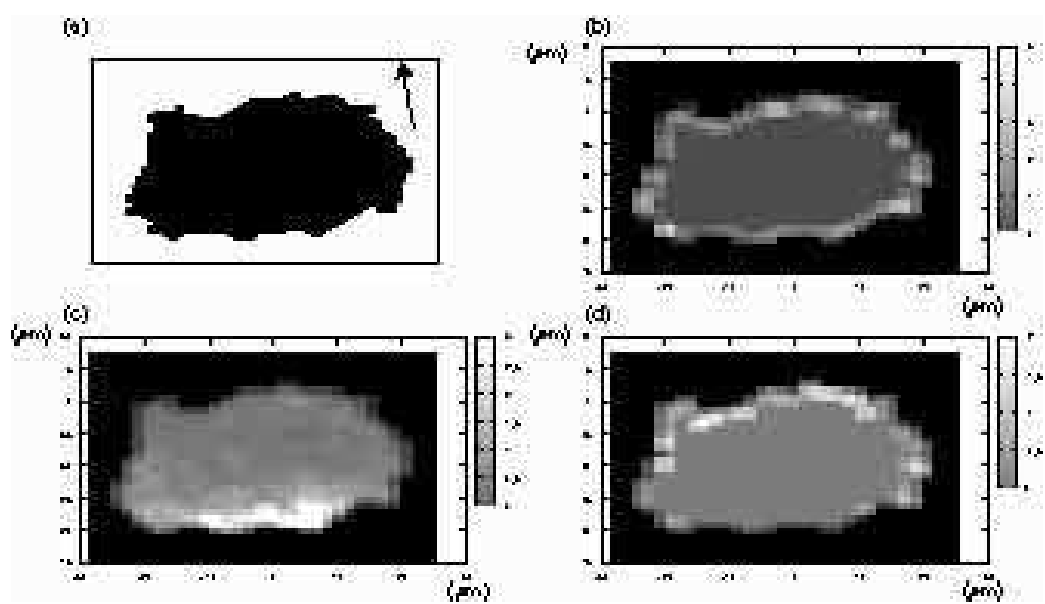}
\caption{(a) A snapshot of the cell shape with $\alpha=1.6$.  
(b) Distribution of activator. (c) Distribution of inhibitor. 
(d) Distribution of actin filaments.}
\label{SNAP2}
\end{center}
\end{figure}

When the parameter $\alpha$ is set to 1.6, the shape becomes 
keratocyte type, that is, the short axis of the cell shape
roughly parallels its moving direction (Fig .  \ref{SNAP2}(a)).
The external signal in Fig. \ref{SNAP2} has the same linear gradient as in Fig. \ref{SNAP1}.
Localization of activator and inhibitor again leads to creation of
actin filaments at the front of cell as shown in 
Fig. \ref{SNAP2} (b), (c) and (d). 
These features agree with the experimental observations of keratocytes \citep{Svitkina97,Verkhovsky99}
and those of the keratocyte-like cells \citep{Asano04}.
Note that the cellular shape with $\alpha=1.0$ or $\alpha=1.6$ in the uniform external signal 
is almost same as that in Fig. \ref{SNAP2}.

In order to analyze the cell shape quantitatively, we calculate 
the direction of "minor  axis"  by fitting an ellipse to the the cell shape;
%
\begin{eqnarray}
(e_x,e_y)  = \left ( \frac{1}{Z} \sum_{i=1}^{N}  x_{i0}y_{i0} ,  
\frac{1}{Z} \left[\sum_{i=1}^{N}y_{i0}^{2} -M \right] \right),
\end{eqnarray}
where $e_x$ and $e_y$ are 
$x$ and $y$-components of a vector pointing along the minor axis and 
$Z$ is the normalization factor to make $e_x^2+e_y^2=1$. $(x_{i0}, y_{i0})$ is 
relative position of the $i$th cellular grid measured  from
the center of mass of the cell. 
$M$ is calculated as
$M= \frac{1}{2}\left(m_{xx}+m_{yy}+\sqrt{(m_{xx}-m_{yy})^2+4m_{xy}^2)}\right)$, 
where $m_{xx} = \sum_{i=1}^{N} x_{i0}^2$, $m_{yy} = \sum_{i=1}^{N} y_{i0}^2$ 
and $m_{xy} = \sum_{i=1}^{N} x_{i0}y_{i0}$.

The cellular velocity $(v_x(t_n),v_y(t_n))$ at time $t_n$ is calculated as \break
$v_x(t_n)=(g_x(t_{n+m})-g_x(t_n))/m\delta t$ and $v_y(t_n)=(g_y(t_{n+m})-g_y(t_n))/m\delta t$, 
where $(g_x(t_n),g_y(t_n))$ indicates center of mass of the cellular domain at 
step $t_n$, and $\delta t$ is a step length in the simulation.  
We select an integer $m$ so that $v(t)$ does not pick up the small fluctuation. 

\begin{figure}[p]
\begin{center}
\includegraphics[scale=0.5]{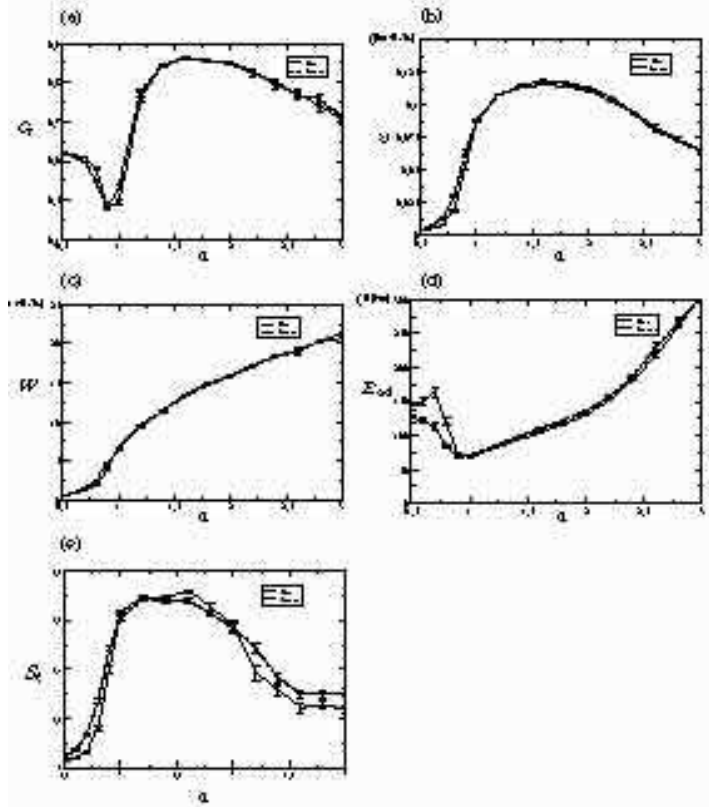}
\caption{Relationships between the parameter $\alpha$ 
and five values, $C_r$, $v$, $W$, $E_{ad}$ and $S_t$.
Solid and dashed lines indicate simulations in the environment of 
the external signal having the linear 
gradient and the uniform concentration, 
respectively. The horizontal axis of all graphs is the 
parameter $\alpha$. (a) The vertical axis is $C_r$: 
correlation between cellular velocity and the semimajor
axis of the cell. Error bars indicates standard deviation. 
All points are  values averaged over 50 different simulations 
started from different random seeds.
(b) $v$: Averaged  speed. (c) $W$: The number of grids 
created by the fourth rule  per unit time, which is 
thought to be proportional to the number of consumed ATPs. 
(d) $E_{ad}$: the number of grids created by the 
fourth rule  per unit length. (e) $S_t$: 
The distance from start point to end point divided by 
the total path length of the trajectory of the center of mass.}
\label{GRAPH}
\end{center}
\end{figure}

Correlation between the axis and the cellular velocity, $C_r$, 
is calculated by $C_r = <|v_x  e_x + v_y e_y |>/\sqrt(v_x^2+v_y^2) >$, 
where the bracket $<>$ indicates average over both the time interval in each simulation run and 
runs started with different random seeds. 
If there is no correlation between velocity and the direction of minor axis, 
$C_r$ should equal to $\frac{1}{2\pi} \int_{0}^{2\pi} \left|{\rm cos} \theta\right| 
d\theta = 2/\pi \sim 0.6366$.  
Therefore, we define that the cell is keratocyte and neutrophil type if 
$C_r$ is significantly larger and smaller than $0.6366$, respectively. 
Fig. \ref{GRAPH}(a) indicates $C_r$ as a function of the parameter 
$\alpha$, which has a minimum around $\alpha=1.0$ and  a 
maximum around $\alpha=1.6$, showing that the cell becomes neutrophil type
around $\alpha=1.0$, and becomes keratocyte type around $\alpha=1.6$.
Note that $C_r$ seems to be closer to $0.6366$ when $\alpha$ is smaller, showing that 
the correlation vanishes between the cellular shape and the velocity direction. This is because 
the cellular shape becomes approximately spherical and the cell fluctuates isotropically as $\alpha$ approaches 0. 

Fig. \ref{GRAPH}(b)  shows the relationship between the averaged cellular 
speed \break \mbox{$v=<\sqrt{v_x^2+v_y^2}>$} and $\alpha$. The peak exists around 
$\alpha=1.6$, located at the same $\alpha$ at which 
the correlation $C_r$ shows the peak, which implies that the keratocyte type cell runs faster than 
cells with other value of $\alpha$.

Fast movement, however, should require more amount of chemical energy.
Since ATP is consumed in actin polymerization, amount of ATP molecules
used or amount of chemical energy required should be proportional to
the number of grids, $N_g$, created by the fourth rule.
Fig. \ref{GRAPH}(c)  shows the relationship between $\alpha$ 
and $W=<\delta N_g/\delta t>$, where $\delta N_g$ is
the number of grids created by the fourth rule per one simulation step. 
$W$ gradually increases as $\alpha$ increases. 
If energy source would be limited and a cell dose not need to 
rush to its destination, it is efficient for the cell to reduce 
``energy per unit length''. We estimate ``energy per unit length'' as follows:
\begin{eqnarray}
E_{ad} = \left< \frac{\sum_t \delta N_g/\delta t}{v(t) T} \right> ,
\end{eqnarray}
where $T$ indicates the time interval of each simulation.
Fig. \ref{GRAPH}(d) indicates that there is a minimum of $E_{ad}$ 
around $\alpha=1.0$, implying that the neutrophil 
type cell moves most efficiently. 

In order to arrive at cell's destination rapidly,  
the cell should move as straight as possible. 
To estimate this straightness we calculate $S_t$ by the distance  
from the start point to the end point divided by the total 
path length of the trajectory of the center of mass of the cellular domain. 
$S_t$ ranges from 0 to 1 by definition.
$S_t$ approaches 1 when the trajectory becomes completely straight. 
Fig. \ref{GRAPH}(e) shows 
that in the uniform distribution of the external signal 
(indicated by a solid line) a peak exists at $\alpha=1.6$, 
whereas in the linear gradient distribution (dashed line) a flat 
peak exists from $\alpha=1.2$ to $1.6$.  
Although there is a slight difference in $S_t$ of keratocyte type ($\alpha=1.6$) 
and $S_t$ of neutrophil type ($\alpha=1.0$), straightness of two types seems almost equivalent.

\section{Discussion}

\begin{table}[h]
\begin{center}
\begin{tabular}{|c|r|r|r|}
\hline
                 & $\alpha=0.4$, $\gamma=100$     &  $\alpha=1.0$, $\gamma=4.0$   & $\alpha=1.6$, $\gamma=4.0$  \\
\hline              
$C_r$       & $0.51\pm0.15$                           &  $0.53\pm0.0065$                      & $0.86\pm0.0030$                    \\
\hline
$v$           & $0.11\pm0.0028$                       &  $0.088\pm0.001$                      & $0.12\pm0.00057$                  \\
\hline
$W$         & $11\pm0.26$                              &  $6.2\pm0.068$                           &  $13\pm0.041$                        \\
\hline
$E_{ad}$ & $97\pm 1.9$                                &  $71\pm0.39 $                              & $110\pm0.51$                         \\
\hline
$S_t$      & $0.48\pm0.035$                           &  $0.61\pm0.0085$                       &$0.68\pm0.0064$                     \\
\hline
\end{tabular}

\end{center}
\caption{}\label{table}

\end{table}%

Our model shows that the keratocyte type cell moves 
more rapidly and more straight than other types of cells. 
These features agree with experimental observations that 
keratocytes \citep{Lee93} or the keratocyte-like $amiB^{-}$ cells \citep{Asano04} crawl 
rapidly along a straight direction. 
In our model this rapid motion is because of the wide spreading
cell membrane at the cell front: Since actin polymerization takes place in the 
wide area in the keratocyte type cell, the driving force of the forward protrusion is stronger than 
in the cell with a narrow leading edge.

\begin{figure}[p]
\begin{center}
\includegraphics[scale=0.5]{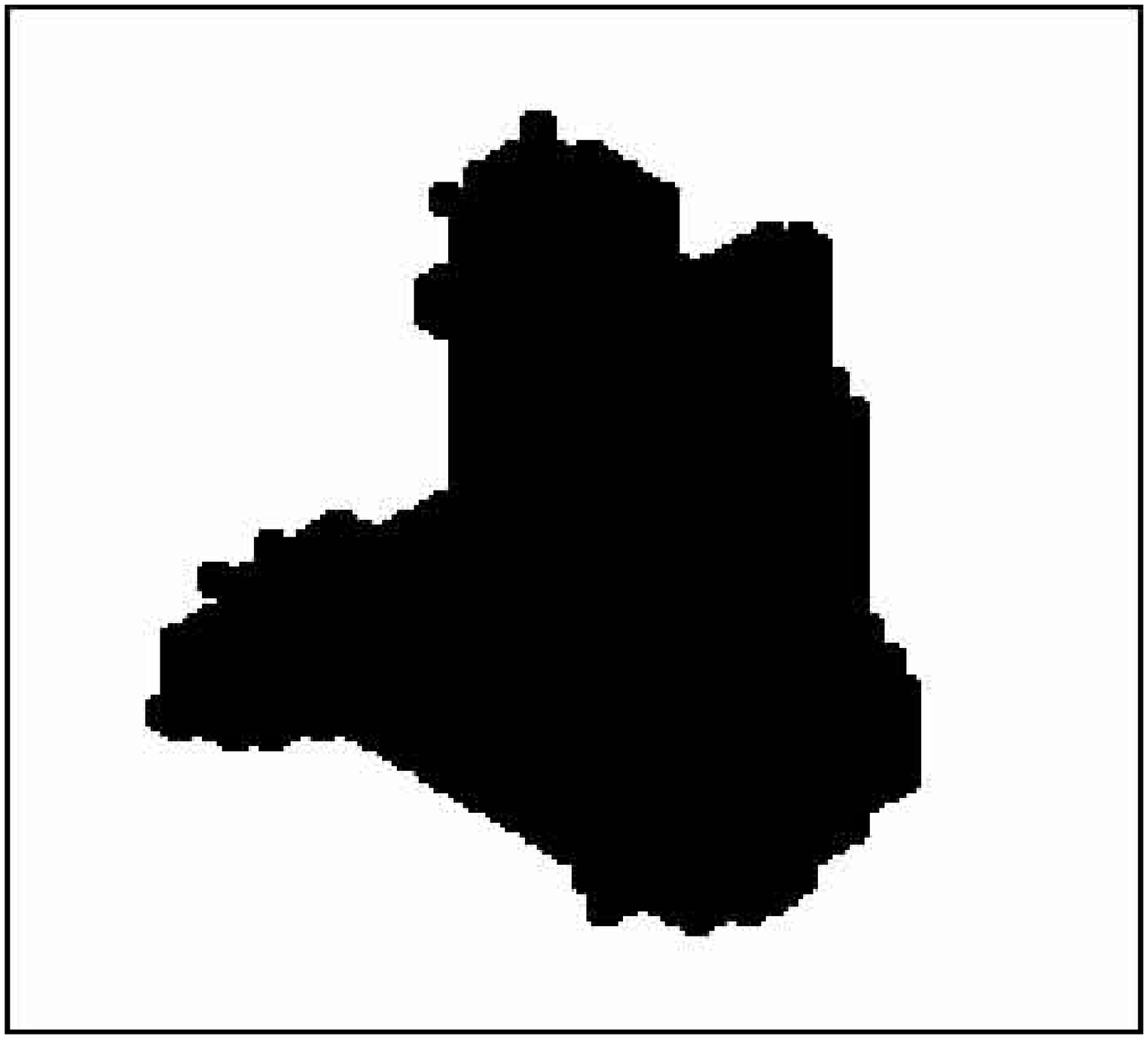}
\caption{A snapshot of the cell with $\alpha=0.4$ and $\gamma=100$.}
\label{snap3}
\end{center}
\end{figure}

Another way to realize the rapid motion is to accelerate the polymerization itself. 
In our model the actin polymerization rate, $\gamma$, controls the rate 
of actin polymerization. We increase the value of $\gamma$ to 100,
meaning that the actin polymerization rate increases twenty-five times. 
This large change of $\gamma$ is consistent with the fact that the rate 
of actin polymerization can sensitively vary tens or a hundred times 
according to the concentrations of G-actin, profin, and other related proteins \citep{Kovar06}.
In order to prevent the cell front from expanding explosively, 
The creation rate of activator, $\alpha$, is suppressed to be 0.4. Table \ref{table} compares $C_r$, $v$, $W$, 
$E_{ad}$ and $S_t$ for three parameter sets with different $\gamma$ and $\alpha$
in the linear gradient condition. The parameter set 
$(\alpha=0.4,\gamma=100)$ allows the cell to move as fast as the keratocyte type
but the motion is less straight. 
The cell with this parameter set is the neutrophil type having $C_r$ value less than $0.6366$;
$C_r=0.51\pm0.15$ as summarized in Table \ref{table}. Its detailed shape, however, is 
different from that in Fig. \ref{SNAP1}(a). 
Large fluctuation of the cell shape as shown in 
Fig. \ref{snap3} is characteristic to this rapidly moving neutrophil type cell, 
which leads to the less straightforward motion of the cell.
Consumed energy is less than the keratocyte type. 
The wild-type $Dictyostelium$ in the starved phase moves rapidly but along the 
fluctuating direction and seems to correspond to this type of cell.

Efficient energy usage of the rapidly moving neutrophil type predicted in the model 
is consistent with the observed decrease in activated NADH in {\it Dictyostelium discodeum} cells:
Amount of activated NADH should reflect the rate of energy consumption, which 
is large in unpolrized cells in the vegetative phage but decrases in highly polarized 
elongated cells in the starved phase \citep{Nomura}.  

Why does a cell with large $\gamma$ and small $\alpha$ have a fluctuated "Dicyosteium"-like shape? 
We can guess the mechanism by following the simulated behavior of cell: 
Small $\alpha$ suppresses actin polymerization, 
but once the actin polymerization is initiated as a fluctuation, 
the large $\gamma$ leads to a rapid protrusion. In this way the large $\gamma$ and 
small $\alpha$ makes protrusion abrupt and intermittent. 
This feature of movement of Dicyosteium-like cell gives a sharp contrast to
that of the keratocyte type cell.
The keratocyte type cell with a larger $\alpha$ but with smaller $\gamma$ 
shows the continuous actin polymerization, leading to the smooth progression of cell.
Thus, the cellular shape and movement decisively depend on $\gamma$ and $\alpha$.
Here, we can summarize 
how the cell shape depends on these two parameters:
Small $\gamma$ and small $\alpha$ lead to the slowly moving neutrophil type, 
small $\gamma$ and large $\alpha$ lead to the keratocyte type,
large $\gamma$ and large $\alpha$ lead to the explosive expansion, and
large $\gamma$ and small $\alpha$ lead to a Dicyosteium-like shape.
It should be interesting to further analyze the morphological change in the
two-parameter space of $\gamma$ and  $\alpha$ in a quantitative way.

We omitted describing detailed molecular processes of contraction at the rear.
Instead, we assumed the rule of keeping cellular size. 
Although our results showed that this assumption explains several 
patterns of amoebae, the assumption clearly does not explain the cell 
locomotion which shows the temporal separation between the protrusion phase 
and the contraction phase \citep{Uchida03}.  
We should improve the model to explicitly take account of the contraction processes, 
which is left as a further avenue of research.

\section{Conclusion}

By using a theoretical model we showed that the rate of production of activator
is a key factor to determine whether the cell shape becomes the neutrophil type or the keratocyte type.
The apparent morphological conversion between two types can be caused by rather simple 
biochemical or genetic perturbations which modulate the activator production rate.
The keratocyte type cells crawl rapidly along the straight direction by consuming a large amount of 
energy. Locomotion of the neutrophil type cell is most economic in terms of energy consumption.
This  slower neutrophil type cell can crawl as straight as the keratocyte type cell. 
The neutrophil type cell can crawl as rapid as the keratocyte type cell when the 
rate of actin polymerization is large enough. The moving direction of 
this rapidly crwaling neutrophil type cell is more fluctuating than the keratocyte type cell.
A plausible explanation of the reason why starved $Dictyostelium$ cells take 
strategy of the rapid neutrophil type locomotion is that they have 
to gather rapidly but the fine tuning of moving direction with fluctuating trials and errors
should be needed when they approach the aggregation center.
Further systems biological investigations through simulations with the model 
which treats the morphological change and chemical reactions in a 
unified way should enable to classify a variety of different strategies of locomotion
and should provide a guide line to design experiments.

\section*{Acknowledgements}

We thank  Drs. Masahiro  Ueda and Hiroaki Takagi for discussions and helpful suggestions. 
This work was supported  by a Grant-in-Aid for the 21st Century COE for Frontiers of Computational Science.

\end{document}